\input epsf.sty
\documentclass{aa2}
\begin{document}

\def\Agata{R\' o\. za\' nska~}

%
%
   \title{Universal spectral shape of high accretion rate AGN}
\titlerunning{Universal spectral shape of AGN}
\authorrunning{Czerny et al.}

   \author{B. Czerny
           \inst{1},
          M. Niko\l ajuk 
          \inst{1},
	A. \Agata
	   \inst{1},
          A.-M. Dumont
         \inst{2},
	  Z. Loska
         \inst{1},
	\and
           P.T. \. Zycki\inst{1}
          }

   \offprints{B. Czerny}

   \institute{$^1$Copernicus Astronomical Center, Bartycka 18, 00-716 
    Warsaw, Poland \\
   $^2$Observatoire de Paris-Meudon, DAEC, Meudon, France \\
             }

   \date{Received ...; accepted ...}


   \abstract{The spectra of quasars and NLS1 
galaxies show
surprising similarity in their spectral shape. They seem to scale only with
the accretion rate. This is in contradiction with the simple expectations from
the standard disk model which predicts lower disk temperature for higher 
black hole mass. Here we consider two mechanisms modifying the disk spectrum:
the irradiation of the outer disk due to the scattering of the flux by the 
extended ionized medium (warm absorber) and the development of the warm
Comptonizing disk skin under the effect of the radiation pressure 
instability. Those
two mechanisms seem to lead to a spectrum which indeed roughly scales, as 
observed, only with the accretion rate. The scenario applies only to objects
with relatively high luminosity to the Eddington luminosity ratio for which
disk evaporation is inefficient.
}

\maketitle
%

\section{Introduction}

The idea of the universal spectral shape of AGN appears in the literature
since many years. The first paper devoted to fitting accretion disk models
to the observed optical/UV spectra stressed that the maximum disk temperature
was always about 20 000 K (Malkan \& Sargent 1982). The stream of data 
from the 
X-ray band seemed at first to reverse the direction of research, as the
variety of the X-ray activity level with respect to the optical/UV was
discovered, together with variety of X-ray spectral slopes.
Nevertheless, the issue of universal shape kept reappearing. 

Walter \& Fink (1993) claim that the spectra of all active galactic nuclei
(AGN) consist of two basic
universal components. The first component is an IR-X ray
power law, with the energy index of the order of 1. The X-ray part of this
component is usually named the ``primary power law'' in X-ray data
analysis (see e.g. Mushotzky et al. 1993). The second
component is the Big Blue Bump, extending from the optical band through
the UV up to the soft X-ray band. Walter \& Fink (1993) argue that objects
differ only in the normalization of the Big Blue Bump with respect to
the IR-X-ray power law, with the shape of the Big Blue Bump being
universal. This view is supported by the statistical analysis of the
$\alpha_{\rm ox}$ parameter measuring the mean spectral slope between
2500~\AA ~ and 2 keV. Such a parameter measures the relative strength
of the Big Blue Bump and roughly represents the overall spectral
shape.  Again, surprisingly, this parameter does not show any strong
trend (Bechtold et al. 2003), and the dispersion is partially due to the
fact that some sources are significantly obscured in the optical/UV band,
like e.g. MCG -6-15-30 (Reynolds et al. 1997).  Also a recent study of
the X-ray spectra of AGN with X-ray luminosity range between $10^{41}$
erg s$^{-1}$ and $10^{47}$ erg s$^{-1}$ did not reveal any trend in the
X-ray slope with
the luminosity (Reeves \& Turner 2000).

Similar surprising uniformity is found in another line of research.
Reverberation studies of the broad line region show that its size is quite
well determined by a {\it single} parameter, in this case the value
of the $\nu L_{\nu} $ flux at 5100 \AA ~(Kaspi et al. 2000). The dispersion 
in this relation is relatively low, less than a factor of 3. 

This broad-band spectral-shape uniformity encouraged the researchers to
connect uniquely the bolometric luminosity of the source with the flux
measured at a single frequency. For example, many authors are using a
simple formula to estimate the bolometric luminosity in a form 
$L = 9 \times \nu L_{\nu} (5100 \AA) $ (e.g. Kaspi et al. 2000), and this
approach seems to reproduce the results of more accurate
integrating over the broad-band spectra quite well (Collin et
al. 2002, Niko\l ajuk et al., in preparation).

This kind of universal picture is by no means expected from the simple 
accretion disk model. The bolometric luminosity,
including X-ray emission, should scale with the accretion rate, $\dot M$,
while the optical flux, coming from the outer parts of the disk, should
scale as $(\dot M M)^{2/3}$ (see e.g. Tripp et al. 1994), 
and thus is additionally influenced by the 
black hole mass, $M$. The soft X-ray slope can in principle take any value,
since apparently there are no constraints for the parameters of the 
Comptonizing plasma.

In the present paper we address the issue of the possible physical 
mechanisms which may account for the universality of the broad-band 
spectra in the case of objects accreting with high accretion rates.

\section{Examples of spectra and corresponding standard disk models}
\label{sec:obsevations}

Radio-quiet quasars and Narrow-line Seyfert 1 galaxies are the sources
expected to accrete mass at a high rate (e.g. Pounds et al. 1995, Collin et al.
2002). Here we choose several extreme examples of such spectra.

\begin{figure}
\vskip -0.5 truecm
\epsfxsize=8.8cm \epsfbox{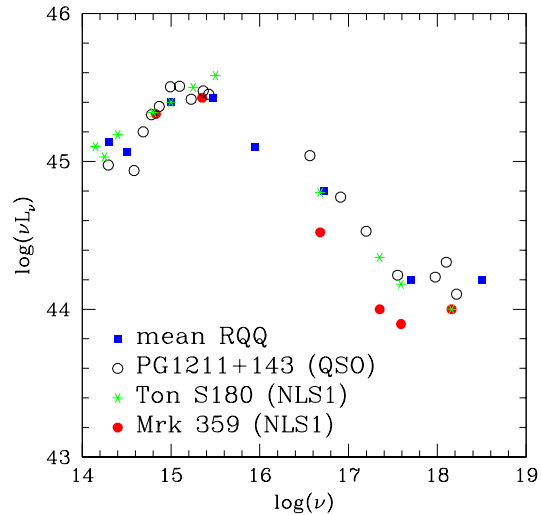}
\vskip -0.8 truecm
\caption{The broad-band composite spectrum of quasars (filled squares) and 
broad-band spectra of PG1211+143 (open circles), Ton S180 (stars) and
Mrk 359 (filled circles), shifted by 0.1, 0.6 and 2.0, respectively,
to coincide at the B band}
\label{fig:examples}
\vskip -0.4 truecm
\end{figure}

A particularly well-covered broad-band spectrum available is the composite
spectrum of quasars (Laor et al. 1997) shown in Fig.~\ref{fig:examples}.
Such a composite seems to be indeed quite representative for all 
high-accretion-rate sources. As examples, we include in the same plot the 
broad-band spectrum of quasar PG1211+143 (Janiuk et al. 2001) and  two 
Narrow-line Seyfert 1 galaxies - Ton S180 (\Agata et al. 2003) and Mrk 359 
(Mathur et al. 2001; O'Brien et al. 2001). 
The bolometric luminosity corresponding to those spectra is 
$1.0 \times 10^{46}$ erg cm$^{-2}$s$^{-1}$,  
$9.0 \times 10^{45}$ erg cm$^{-2}$s$^{-1}$,  
$2.5 \times 10^{45}$ erg cm$^{-2}$s$^{-1}$ and 
$9.7 \times 10^{43}$ erg cm$^{-2}$s$^{-1}$,
assuming a Hubble constant equal to 50 km s$^{-1}$ Mpc$^{-1}$.
The optical flux for Mrk 359 has been corrected for starlight contribution (factor
2.5) on the basis of the ratio of the observed value of the $H_{\beta}$ flux
in the starlight-free source Ton S180 (46 \AA) and Mrk 359 (18 \AA) 
(V\'eron-Cetty et al. 2001). 

The sources covering two orders of magnitude in bolometric luminosity, and
therefore possibly two orders of magnitude in mass, show nevertheless
a surprisingly similar broad-band shape. The description of the X-ray spectra of
Ton S180 and  Mrk 359 may seem different (the X-ray photon index of Ton S180 is
given as 2.44, Turner et al. 2001 and of Mrk 359 as 1.85, 
O'Brien et al. 2001). However, this
apparent difference mostly reflects the fact that the instruments used are 
most sensitive at 2-3 keV and return, as 'representative' X-ray slope, the
slope around this energy. If for example, we fit the Ton S180 Chandra data
of Turner et al. (2001)
with a broken power law model ({\sc bknpower} in XSPEC software) and fix the 
second slope at 1.85 we obtain a 
satisfactory fit, with the energy break at $2.7^{+0.2}_{-0.1}$ keV 
instead of at 2.0 keV,
as apparently seen in the XMM data of Mrk 359. The two spectra seem to be shifted
in energy only by a small factor of 1.35, or 0.13 in the logarithm!

The bolometric luminosity is dominated by the optical/UV/far-UV part
($\sim 70 $ \%), the EUV-soft X-ray band contributes about 25 \%, and
a few per cent comes from the hard X-ray power law.

\section{The model}

It is well known that the simplest standard Shakura-Sunyaev disk model
(Shakura \& Sunyaev 1973), under assumption of local black-body
emission, cannot account for the observed X-ray spectra of AGN (Czerny
\& Elvis 1987; Mushotzky et al. 1993; Krolik 1999; for a recent
review, see e.g. Czerny 2002). The model is also not quite
satisfactory in the optical/UV band since the observed spectra are
frequently redder than predicted by the model. The model spectra, in the long
wavelength limit, have the universal slope of 1/3 while the mean
quasar slope above 1050 \AA~ in the Zheng et al. (1997) sample is
$-0.99 \pm 0.05$.

The problems with the model are progressively more serious with a
decreasing Eddington ratio. There are
several arguments in favour of the view that in low luminosity sources,
like normal Seyfert 1 galaxies, the standard disk exists only in the
outer part of the flow while closer to the black hole it is replaced by
the inner hot optically thin flow, possibly ADAF(e.g. Ichimaru 1977;
Narayan \& Yi 1994; Abramowicz et al. 2002). We meet a similar situation
in the galactic black holes in their hard/low luminosity state
(e.g. Esin et al. 1997; Nowak et al. 2002).  Modeling of the flow
requires a deep change in the basic geometry, in comparison with the
standard model. There were ideas of the optically thick (but highly
ionized at the surface) disks dominating the flow even in the
innermost part (e.g. Beloborodov 1999; Poutanen \& Fabian 1999) but
a recent study of the hard X-ray tail of the reflection component in Cyg
X-1 suggests that the disk is indeed disrupted, and not just invisible
due to high reflectivity (Barrio et al. 2002).

However, for sources with relatively high Eddington
ratio, like quasars (see e.g. Collin et al. 2002) and
Narrow-line Seyfert 1 galaxies, there are arguments in favour of the
standard disk extending down to the marginally stable
orbit. The reflection component in those sources is significantly ionized
but strong. Relativistic smearing indicates that this component comes from the
innermost part of the flow (e.g. Janiuk et al. 2001).  Also the iron
line is strong and (relativistically) broadened in many of those
sources (MCG-6-30-15, Lee et al. 2002; Ark 564, Comastri et al. 2001;
low redshift quasars, Mineo et al. 2000).  In those sources (see
Fig.~\ref{fig:examples}) the bolometric luminosity is dominated by the
Big Blue Bump which can be roughly explained as coming from the disk.

\begin{figure}
\vskip -0.5 truecm
\epsfxsize=8.8cm \epsfbox{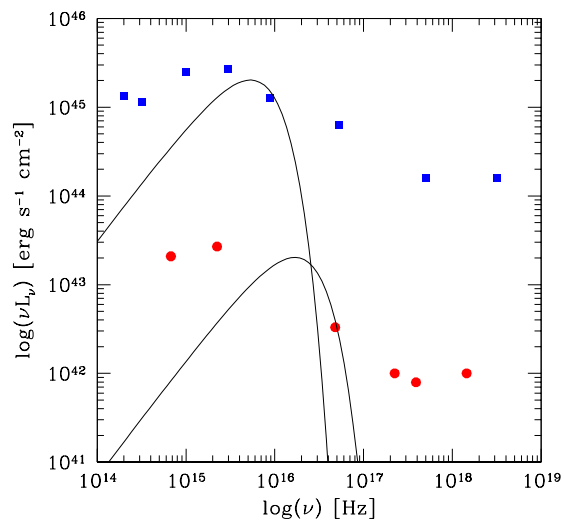}
\vskip -0.8 truecm
\caption{Points show two examples of AGN spectra: the composite 
spectrum of quasars (filled squares) and Mrk 359 (filled circles). 
Standard accretion disk model spectra for two values of
a black hole mass ($ 10^6
M_{\odot}$ and  $10^8
M_{\odot}$) are plotted with 
continuous lines. In both 
cases we assumed the Eddington ratio $L/L_{\rm Edd}$ equal 0.3. 
}
\label{fig:standard}
\vskip -0.5 truecm
\end{figure}

Therefore, in the case of high-luminosity sources, all we need is some
modification of the standard model, and such modifications were
introduced in the literature whenever the model was compared with the
observational data.  X-ray spectra were modeled by adding some kind of
a corona or warm skin to the disk (Czerny \& Elvis 1987; Shimura \&
Takahara 1995; Kawaguchi et al. 2001).  Optical/UV spectra were
modeled by introducing an irradiation of the outer disk by the
innermost part (e.g. Czerny et al. 1994, Loska \& Czerny 1997, Soria
\& Puchnarewicz 2002). However, those modifications were basically
arbitrary from the point of view of our knowledge of the accretion
process, and therefore do not explain the reason why the resulting
spectral shape is so universal. Comparison of the data with the
simplest standard disk models (see Fig~\ref{fig:standard}) shows
a significant trend with the black hole mass. It also indicates what 
fraction of the energy has to be
reprocessed in order to explain the observed spectral shape.

In the two subsections below we propose two modifications to the
standard model which directly result from our knowledge of
accretion disk physics and AGN surroundings.

\subsection{Irradiation and the optical/UV spectra}

The irradiation of the outer parts of the disk may be either direct or
indirect, through scattering the emission from the innermost part by
some medium and redirecting  part of the scattered flux towards the
outer disk.

\subsubsection{Direct irradiation}

Direct irradiation was discussed by several authors in the context
of AGN (e.g. Rokaki et al. 1993; Loska \& Czerny 1997; Nayakshin \&
Kazanas 2002) or GBH (e.g. Cannizzo 2000), mostly concentrating on
illumination of the disk by an X-ray source, arbitrarily located high
above the disk. Since in high $L/L_{\rm Edd}$ sources the hard X-ray flux
is small this effect is not as important as the illumination by the
inner disk emission. 

Self-irradiation is efficient if the disk is
strongly flaring, since the direct irradiation by the point-like source is
roughly given by $ F_{direct} =
\gamma L / (4 \pi R^2)$ (see e.g. King 1998, Soria \& Puchnarewicz
2002), where $\gamma$ is determined by the disk thickness, $H$, as
$\gamma = (H/R)(d \ln H / d \ln R - 1)$.  We can roughly estimate the
effect at the distance of a few thousands of Schwarzschild radii from
the Shakura-Sunyaev paper, their region (b) or (c). The value $\gamma$
is of order of $10^{-4}$, since it is systematically smaller in AGN
than in GBH due to the direct scaling with mass seen in the formula
for the disk thickness. The same conclusion was reached in the
paper of Kurpiewski et al. (1997). 

Additionally, if we actually consider
not a point-like isotropic source but an inner flat disk, with
effective radiation flux proportional to the inclination angle measured with
respect to the symmetry axis,
the appropriate expression for $\gamma$ is $\gamma = 2 (H/R)^2
(d \ln H / d \ln R - 1)$ (see King et al. 1997 and references therein). 
The presence of an additional factor $H/R$ again lowers the effectiveness of 
direct self-irradiation. 

Numerical computations, including a more
accurate approach to the disk opacity (bound-free transitions),
support this results (Loska et al.,
in preparation). Due to the complex dependence of the disk thickness
on the distance only a relatively narrow belt at intermediate radii
as well as very distant parts of the disk are irradiated directly
by the disk central regions. Direct irradiation was
efficient in the paper of Soria \& Puchnarewicz (2002) since they
adopted a much higher value of $\gamma$ (of order of 0.03) than 
expected for an AGN.

Therefore, in the present paper we neglect the effect of direct 
irradiation.

\subsubsection{Indirect irradiation}

The indirect irradiation was studied in detail in a number of papers
devoted to the accretion disk coronae (e.g. Ostriker et al. 1991;
Murray et al. 1994; Kurpiewski et al. 1997; see also Esin et al. 1997
for the case of galactic sources). The effect was found to be
important if the corona is of considerable optical depth and if it is
geometrically thick. However, the existence of a corona with the
appropriate parameters is neither obvious, nor widely
accepted. Attempts to reproduce the corona on the physically sound
basis of disk evaporation (e.g. Meyer \& Meyer-Hofmeister 2000; \Agata
\& Czerny 2000; Liu et al. 2002) predict so far a very weak and
optically thin corona in the case of sources with even moderately high
$L/L_{\rm Edd}$.

In the present paper we propose a different mechanism for the indirect
disk irradiation, namely scattering by the warm absorber.

\subsubsection{Warm absorber properties}

The warm absorber is a significantly ionized medium present in our
line of sight towards sources classified as Seyfert 1 galaxies, Narrow
Line Seyfert 1 galaxies or quasars.  Observationally, it is seen
through the presence of narrow absorption lines and absorption
continua in the soft X-ray band which are well resolved now in the
Chandra or XMM data (see Kaastra et al. 2002; Lee et al. 2001;
Steenbrugge et al. 2003; Sabra et al. 2003).  Most lines come from
highly ionized species like CIV, OVII and OVIII, suggesting that the
temperature of the medium is of the order of $10^5 - 10^6 $ K. The column density
of the warm absorber is frequently of the order of $10^{23}$ cm$^{-2}$ or
more, and the medium is extended but probably clumpy (see e.g. Krolik
2002).

Although we see the presence of the warm absorber due to absorption
features superimposed onto the transmitted spectrum from the nucleus
this medium actually predominantly scatters the incoming photons
instead of absorbing them. Most of the radiation from the nucleus is
emitted in the far UV band while at such wavelengths the electron
scattering dominates the total cross-section.

\begin{figure}
\epsfxsize=8.8cm \epsfbox{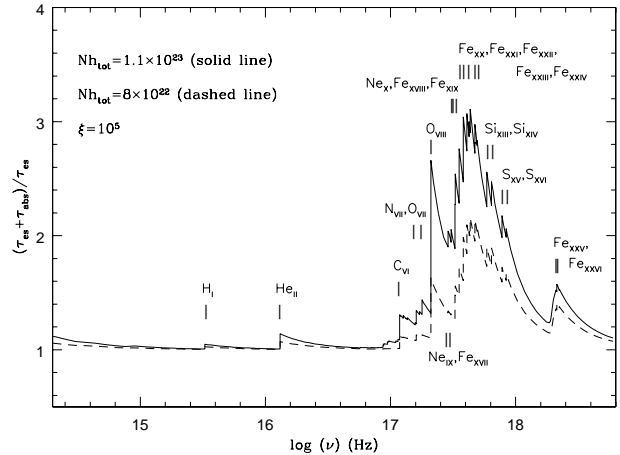}
\caption{The ratio of the total optical depth 
(absorption plus scattering) to the Thomson optical depth (scattering)
as a function of energy for the warm absorber model in Ton S180:
ionization parameter at the surface $\xi = 10^5$, incident flux with
photon index 2.6 extending from 12 eV to 100 keV, density profile
from the constant pressure condition, total column density 
$1.1 \times 10^{23} $ cm$^{-2}$ (solid line) and $8 \times 10^{22} $ cm$^{-2}$
(dashed line). }
\label{fig:warm}
\end{figure}

We illustrate the scattering efficiency of the warm absorber in
Fig.~\ref{fig:warm}.  It shows the ratio of the total optical depth of
the warm absorber to the Thomson optical depth as a function of the
energy, calculated for an exemplary case of a warm absorber model
considered as appropriate for Ton S180 (\Agata et al. 2003).  The
model is computed with the use of the radiation transfer code {\sc
titan} developed by Dumont et al. (2000) and modified for the purpose
of warm absorber modeling by \Agata et al. (2003).
We assumed constant pressure throughout the medium, as advocated by 
Krolik (2002), since it may better represent the density profile within 
separate clumps. 

The emission spectra of the considered objects (see
Sect.~\ref{sec:obsevations}) peak in the frequency range
$10^{15} - 10^{17}$ Hz where electron scattering dominates.

\begin{figure}
\epsfxsize=7.0cm \epsfbox{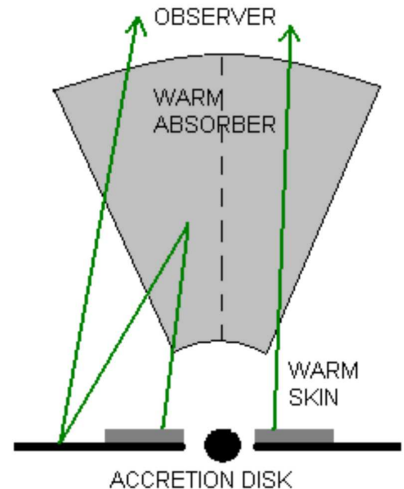}
\vskip -0.3 truecm
\caption{The schematic picture of a bright AGN with a warm absorber and a warm 
disk skin.}
\label{fig:view}
\vskip -0.5 truecm
\end{figure}

\subsubsection{Modification of the disk spectrum due to fully 
ionized warm absorber}

The effect of scattering slightly decreases the overall normalization
of the spectrum, by a factor of $\exp (-\tau_{wa})$, and
redirects a fraction of the scattered radiation back towards the
disk. In the further consideration the effect of absorption is neglected
and only the effect of scattering is considered.

The radial extension of the warm absorber is not well constrained and
distances from light days - weeks to tens of parsecs were proposed
(see Krolik 2002).  We envision the geometry in
Fig.~\ref{fig:view}. We neglect the clumpiness of the medium and we
parameterize the region by the total optical depth for electron
scattering, $\tau_{wa}$, minimum $z_{\rm min}$ and maximum $z_{\rm
max}$ distance of the warm absorber from the nucleus. We assume that
the dependence of the density of the warm absorber on the distance $z$
can be described as a power law, $\rho(z) \propto z^{-\delta}$, with
the normalization fixed by the assumption of the total optical
depth. For our purpose of a rough estimate of the effect we do not
introduce any specific dependence on the distance from the symmetry
axis but we assume that the irradiation of the disk at a distance r
from the center can be approximately represented as
\begin{equation}
  F_{irrad} = { 2 \tau_{wa} f_{wa}(\delta - 1) L \over z_{min}^{1 - \delta} - z_{max}^{1 - \delta}} {\int_{z_{min}}^{z_{max}} {z^{-\delta + 1} \over (z^2 + r^2)^{3/2}} dz}
\label{eq:irrad}
\end{equation}
where $L$ is the bolometric luminosity of the accretion disk and 
the coefficient $f_{wa}$ corresponds to the fraction of nuclear
emission passing through the warm absorber, determined by the solid
angle. The factor 2 is due to the disk emission anisotropy - the flat
central parts of the accretion disk are well approximated by a
central point-like source with luminosity 
$L(i) = 2 L \cos i$ dependent on the inclination angle $i$ measured with 
respect to the symmetry axis (see King et al. 1997 and references 
therein). Since the warm absorber is located predominantly along the 
symmetry axis, $ \cos i \approx 1$. We assume that the irradiation flux 
is subsequently thermalised
within the disk, with albedo equal to 0.2, appropriate for the weakly
ionized surface of an outer disk.

\begin{figure}
\epsfxsize=8.8cm \epsfbox{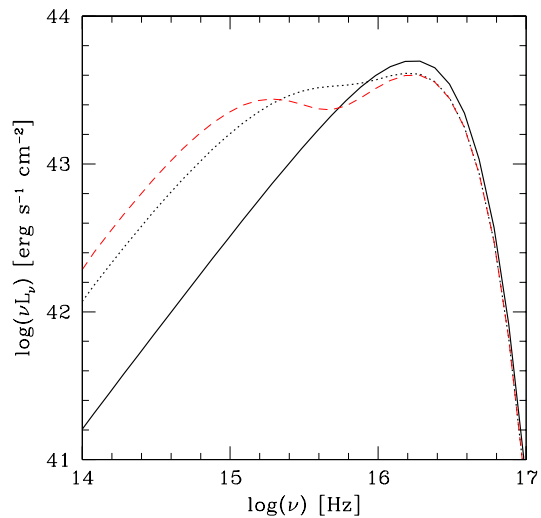}
\caption{The effect of the warm absorber on the spectra of the standard 
disk due to extinction (scattering) and 
illumination by the scattered radiation. Continuous line: disk 
without warm absorber, dotted line: $z_{min} = 100 R_{Schw}$, 
dashed line: $z_{min} = 300~ R_{Schw}$. Fixed parameters: black hole mass 
$1.5 \times 10^6 M_{\odot},$, $L/L_{Edd} = 0.5$, $z_{max} = 5 000~ R_{Schw}$,
$\delta = 2$, $\tau_{wa} = 0.2$.}
\label{fig:irradiation}
\end{figure}

For the range of disk radii of the order of the spatial extension of the
warm absorber such an irradiation decreases more slowly with the disk
radius than the energy dissipation, which is proportional to $r^{-3}$.
The outer disk temperature is enhanced and the disk spectrum is
modified predominantly at longer wavelengths (optical and the IR
band). The effect is only weakly influenced by the location of the
outer edge of the warm absorber and the density distribution, more
important is the exact location of the inner edge of the warm absorber
(see Fig.~\ref{fig:irradiation}).

Since the effect of irradiation depends mostly on the inner disk
bolometric luminosity the resulting spectrum mostly depends on the
bolometric luminosity of the source.  The temperature profile in the
intermediate parts approaches $r^{-1/2}$ distribution, and the spectrum
integrated over the disk surface, $F_{\nu}$, scales roughly as
$F_{\nu} \propto L \nu^{-1}$. The exact shape computed from Eq.~(\ref{eq:irrad}) 
is somewhat more complex than this simple law but it basically preserves 
the scaling with the bolometric luminosity.

\subsection{Stabilization of the disk by a warm skin and formation of the 
soft X-ray spectra}

Comptonization of a fraction of disk photons is clearly needed in
order to reproduce the soft X-ray part of the spectrum. It is not
clear what is the geometry of the Comptonizing medium. 
However, in
the case of high-luminosity objects the favoured model is that of a
disk extending down to the marginally stable orbit and of a warm skin
covering the inner region. A similar interpretation holds both for
galactic black holes in their High State or Very High State and for
luminous AGN (e.g. Czerny 2002; Miller et al. 2002; Gierli\' nski et
al. 2001). The support comes from the high disk temperature
(Puchnarewicz et al. 2001; Kubota et al. 2001), the
presence of broad $K_{\alpha}$ line (MCG-6-30-15, Tanaka et al. 1995;
Lee et al. 2002) and relativistically broadened reflection by the
ionized material (e.g. PG 1211+143, Janiuk et al. 2001).  The hard
X-ray power law may come from separate coronal flares, or
possibly from a contribution of non-thermal electrons to a mostly
thermal population of warm skin electrons, as suggested by Vaughan et
al. (2002) on the basis of the Ton S180 data analysis. Such a view is
observationally supported by the data for Cyg X-1 in its high/soft
state where the hard X-ray tail is faint but extends up to $\sim 10 $
MeV (Gierli\' nski et al. 1999).
 
The properties of the Comptonizing material are not specified from
a theoretical point of view. They are usually adopted as model
parameters to be fitted to the observational data (see e.g. Vaughan et
al. 2002). In the present section we propose a mechanism which leads
in a natural way to a certain value of the Compton amplification factor, 
$A$, of the
warm skin, and consequently, similar soft X-ray slopes in many
objects. We neglect the issue of the non-thermal fraction of the electron
population and the hard X-ray tail of the spectra.

\subsubsection{General stationary model of the warm skin}

We assume that the accretion disk is divided into two layers: an inner
cold part and a warm Comptonizing skin. This is a typical assumption
made in many disk/corona models (e.g. Haardt \& Maraschi 1991; Misra
\& Taam 2002). We further assume that a fraction of the energy is
transported from the disk interior towards the skin as a magnetic
flux. Therefore, the energy balance of the disk interior reads:
\begin{equation}
Q^{-} + W = Q^{+},
\label{eq:balance}
\end{equation}
where $Q^{-}$ is the radiative cooling of the disk interior, $ W$ is
the magnetically transported flux and $Q^{+}$ is the energy flux
dissipated in the disk interior.

We describe the radiative cooling of the optically thick disk and the
dissipative heating in the disk interior as in the basic model of
Shakura \& Sunyaev (1973) with the vertically averaged approach
\begin{equation}
Q^{-} = {4 a T^4 c \over 3 \Sigma \kappa_{es}},
\end{equation}
where $T$ is the disk interior temperature, $\Sigma$ is the 
disk surface density, $\kappa_{es}$ is the 
opacity coefficient for electron scattering and $a$ is related 
to the Stefan-Boltzman constant 
$\sigma_B$ through $\sigma_B = a c/4$;
\begin{equation}
Q^{+} = \alpha P H \Omega_K,
\end{equation}
where $\alpha$ is the standard viscosity coefficient, $P$ is the pressure and 
$\Omega_K$ is the Keplerian angular velocity.

\begin{figure}
\epsfxsize=8.8cm \epsfbox{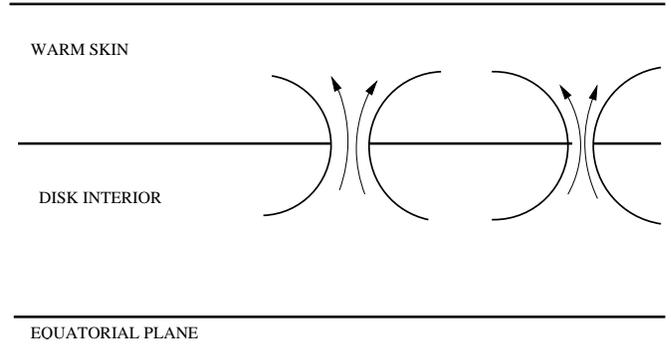}
\caption{The schematic view of the heating of the warm skin.}
\label{fig:skin}
\end{figure}

We now assume that the magnetically transported flux is proportional to the 
energy density of the magnetic field, $B^2$ (which is in turn proportional 
to the 
total pressure with the disk), to the Alfven velocity $v_A$, and to the 
fraction of the surface developing active regions, which we assume to be 
proportional to the disk thickness, $H$, 
\begin{equation}
  W \propto B^2 v_A H.
\label{eq:mag_cooling}
\end{equation}
We illustrate the idea of localized magnetic tubes in Fig.~\ref{fig:skin}.

The first two factors in Eq.~(\ref{eq:mag_cooling}) 
arise in a natural way from considerations of
magnetic activity in the context of a disk corona (e.g. Heyvaerts \&
Priest 1989; Merloni \& Fabian 2002) or the formation of stellar
coronae (e.g. Fawzy et al. 2002). The third factor is more difficult
to estimate on the basis of theoretical considerations. Flux tubes at
any moment usually cover only a fraction of the boundary surface
between the deep interior and the upper stellar coronae but this
covering factor is frequently introduced as a free parameter, 
of the order of 0.1 (e.g. Fawzy et al. 2002). We assume that this covering
is proportional to the disk thickness, and our argument is the
following.

The size of any convective cell in the disk interior scales with the
disk thickness, as argued by Shakura \& Sunyaev (1973), with the
viscosity parameter providing the proportionality constant. A similar
approach was used in the corona model by Galeev et al. (1979). Recent
3-D numerical simulations of the development of the magnetorotational
instability also support such a view (e.g. Miller \& Stone
2000). Therefore, disk thickness defines in a natural way the radial
size of the active region at the border between the inner disk and a
warm skin. In the azimuthal direction, the motion is much faster than
the characteristic convective motion (the sound speed is much smaller than
the Keplerian velocity) so the structures become immediately elongated
and their azimuthal sizes scale with radius and do not depend on the
disk thickness. The number of active zones may be constrained if their
number is related to the observed hard X-ray emission flare. The variability
of the hard X-ray component was studied in a number of galactic sources
and AGN using power spectra analysis. The basic and general result
is that the power spectra and rms (root mean square) variability
depend on the luminosity but when renormalized (i.e. if rms and
power spectrum are divided by the mean luminosity and its square, 
correspondingly) they
become luminosity-independent.  This fact, stressed recently by Uttley
et al. (2002) means that the number of active regions directly
responsible for the observed variations is independent of the
luminosity.  If the number of active regions is constant, their
azimuthal extension is independent of the disk thickness and the radial
size is proportional to the disk thickness. Therefore,  we can assume that
effectively the active fraction of the surface between the inner disk
and the warm skin is proportional to the disk thickness.

The energy density of the magnetic field in the disk interior is
expected to be proportional to the pressure, $P$, i.e.  $B^2 \propto
P$, and the Alfven speed can be expressed as a function of the disk
surface density, $\Sigma$ and the disk thickness, $H$, $v_A = B
H^{1/2}/\Sigma^{1/2}$.

Therefore we obtain the following characteristic dependence
\begin{equation}
W \propto P^{3/2} H^{3/2}.
\end{equation}

\subsubsection{Thermal stability of the disk interior}

We perform next the stability analysis of the disk interior in the
thermal timescale, closely following the analysis done by Czerny
et al. (1987).  We assume that in such a short timescale
the surface density remains constant throughout perturbation. Since the
dynamical timescale is in turn much shorter than the thermal timescale
we assume that the expanding/contracting disk adjusts its structure to
preserve hydrostatic equilibrium. The hydrostatic equilibrium
equation reads
\begin{equation} 
{k \over m_H } {\Sigma \over H} T + {4 \over 3} a T^4 = \Omega_k^2 \Sigma H,
\label{eq:hydro}
\end{equation}
where $m_H$ is the hydrogen mass, and
$k$ is the Boltzman constant. The first term of 
Eq.~(\ref{eq:hydro}) represents the gas pressure and the second term the 
radiation pressure. Perturbing this equation for $\Sigma = const$ gives us the
relation between the change of the disk thickness and  
the perturbation of the disk temperature. It can be most conveniently expressed as
\begin{equation}
 {\delta H \over H} = {4 - 3 \beta \over 1 + \beta } {\delta T \over T},
\label{eq:delta_H}
\end{equation}
where $\beta$ is the gas-to-total-pressure ratio.
Using Eq.~(\ref{eq:delta_H}) we can express the perturbed energy transport terms
as functions of $\delta T$: 
\begin{equation}
\delta Q^+ = 2{4 - 3 \beta \over 1 + \beta }Q^+ \delta \ln T,
\end{equation}
\begin{equation}
\delta Q^{-} = 4 Q^{-} \delta \ln T,
\end{equation}
\begin{equation}
\delta W = 3{4 - 3 \beta \over 1 + \beta} W \delta \ln T.
\end{equation}

The disk is thermally stable if the increase in the heating rate is smaller
than the increase of the cooling rate for positive temperature 
perturbation (e.g. Piran 1978), i.e.
\begin{equation}
  \delta Q^+ < \delta Q^{-} + \delta W.
\label{eq:stab1}
\end{equation}
Introducing a convenient parameter $w$ measuring the importance of the
magnetically transported flux
\begin{equation}
   w= W/Q^{+}
\end{equation}
and using the equilibrium condition given by Eq.~(\ref{eq:balance}) 
we can express the condition given by Eq.~(\ref{eq:stab1}) as
\begin{equation}
\beta > {2 - 4 \, w \over 5 -6.5 \, w}.
\label{eq:stab_criterion} 
\end{equation}
For $w=0$ the criterion reduces to the standard criterion
of Shakura \& Sunyaev (1976).

The presence of magnetic transport stabilizes the disk since its
dependence on the temperature is steep enough (the same as in the case
of an advection term).  When $w = 0.5$ the disk already becomes stable
even if strongly dominated by the radiation pressure 
(i.e. $\beta = 0$).

\subsubsection{Marginal stability solution for the warm skin}

We therefore consider it possible that the disk in a natural way develops
this magnetically mediated transport in the radiation-pressure-dominated 
region and that the transport efficiency saturates at the value $w$ 
corresponding to a marginal stability, 
\begin{equation}
w = 0.5. 
\end{equation}

This region of the disk would be therefore covered with a warm skin. We
named it 'warm', and not a 'hot' skin since we have in mind a medium with the
temperature well below the typical temperature 100 keV invoked for the hot 
Comptonizing medium responsible for the hard X-ray power law with the energy 
index $\sim 0.9$ seen in most (perhaps all) objects instead of, or in 
addition to, the much steeper soft X-ray power law which we intend to model.

\begin{figure}
\epsfxsize=8.8cm \epsfbox{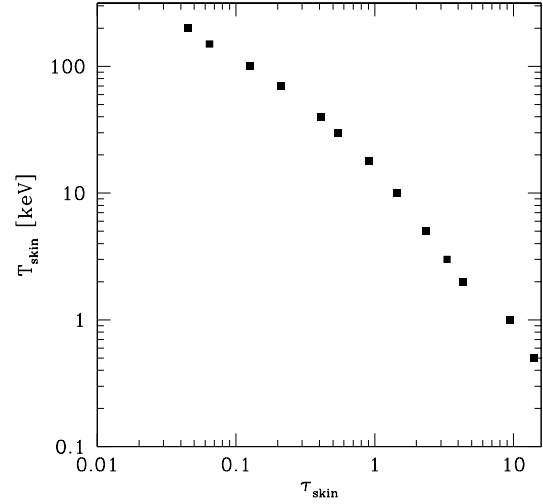}
\caption{The relation between the electron temperature of the warm skin, 
$T_{skin}$, and the optical depth, $\tau$, for a slab with fixed
Compton amplification factor $A=2$ and input photon temperature 5 eV.}
\label{fig:ampli}
\end{figure}

The warm skin cools predominantly by Comptonization, as suggested by
the roughly power law character of the spectrum in the soft X-ray
band. We further assume that the effect of bremsstrahlung 
and synchrotron emission are negligible. We also assume that the disk
surface layers underlying the warm skin are strongly ionized, i.e. the
albedo at the disk/skin division surface is equal to 1. 

The cooling efficiency of the Comptonization is therefore fixed by the
parameter $w$, i.e. the Compton amplification factor, $A$, is given by
\begin{equation}
  A= {Q^{-} + W \over Q^{-}} ={1 \over 1 - w} = 2.
\end{equation}

Our model therefore roughly corresponds to the original Haardt \&
Maraschi (1991) analytical model with the fraction of the energy
dissipated in the corona $f=0.5$ and the albedo $a=1$. However, we
provide a physical justification for the specific choice of $f$ in the
case of the underlying disk dominated by radiation pressure.

The obtained value of the Compton amplification factor, $A = 2$, for 
the warm skin of a marginally
stable disk interior, $w = 0.5$,  
is significantly
lower than the value $ A \sim 3$ determined from the model of Haardt
\& Maraschi (1991) for the corona-dominated solution ($f=1, \ a=0$ in their
Eq.~(3b)). 

Since the Compton amplification factor is related to the 
Compton parameter $y$, it roughly determines the slope of the radiation 
spectrum, $\alpha_E$, through simple analytical formulae 
(see Rybicki \& Lightman 1979).  Therefore, 
 warm skin spectra are expected to be much softer than the
$f = 1$ corona of Haardt \& Maraschi (1991). For example, an approximate
analytical formula of Beloborodov (1999): 
$\alpha_E = 2.33 (A-1)^{-1/10}-1$ gives the expected slope 
$\alpha_E = 1.33$ for $A=2$ and $\alpha_E = 1.17$ for $A=3$ . 
However, analytical formulae are 
not accurate; quantitative results can be conveniently obtained 
from Monte Carlo computations.

\subsubsection{Radiation spectrum from disk/warm skin system}

We perform Monte Carlo simulations assuming plane parallel geometry.
We developed our code from the code used by Janiuk et al. (2000). 
The slab
is parameterized by the electron temperature, $T_{skin}$, and the
optical depth, $\tau_{skin}$. Photons are emitted isotropically at the
bottom of the slab. The spectral distribution of the injected photons
is described either by a black body distribution or provided in a 
numerical form. 
Since the disk is highly ionized we assume a perfect
mirror at the bottom of the slab. Photons trying to leave the slab
through the bottom surface are collected and returned to the medium
with their energy unchanged and their velocity reversed.
This means we neglect, for simplicity, the energy losses during
the scattering of the photons with the bottom surface which in principle
is important for higher energy photons (above $\sim $ 15 keV) due to
the inelastic scattering.   
All photons are finally collected at the slab upper surface. 
The final spectrum 
is obtained by summation of all photons over the angle at
which they leave the surface, i.e.\ the spectrum represents the average over
all inclination angles. 

We perform computations for both
the optically thin and optically thick warm skin. We compute the
amplification factor for a set of models. For each optical depth we
find a value of the temperature corresponding to the amplification
factor equal to 2. The resulting plot is shown in Fig.~\ref{fig:ampli}.

\begin{figure}
\epsfxsize=8.8cm \epsfbox{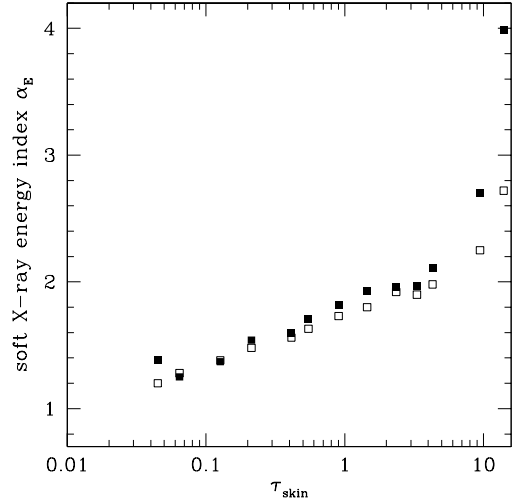}
\caption{The soft X-ray slope of the radiation Comptonized by a slab
with fixed Compton amplification factor $A=2$ and soft photon temperature 5 eV 
as a function of
 the optical depth
of the slab.  Filled squares: slope measured between 0.5 and 2 keV; 
open squares: slope measured between 0.1 and 2.4 keV.}
\label{fig:slope}
\end{figure}

\begin{figure}
\epsfxsize=8.8cm \epsfbox{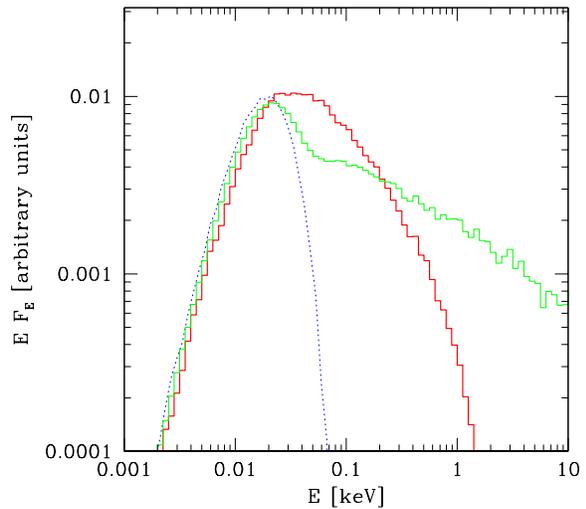}
\caption{The Comptonization of a 5 eV input black body emission (dotted line) 
by a slab
with temperature, $T_{skin}$ equal 0.5 keV 
(thick continuous line) and 100 keV 
(thin continuous line). For both slabs the
Compton amplification factor $A=2$ determines the Compton optical
depth $\tau_{skin}$ (equal 13.98 and 0.127 correspondingly). 
Typical pairs of temperature and
optical depth from Fig.~\ref{fig:ampli} give spectra intermediate 
between these two extreme examples.}
\label{fig:mc_model}
\end{figure}

\begin{figure}
\epsfxsize=8.8cm \epsfbox{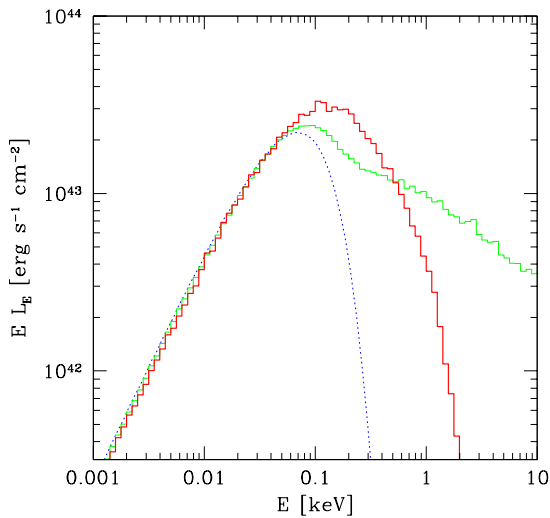}
\caption{The Comptonization of a disk emission instead of a single black body 
for the same two slabs as in Fig.~\ref{fig:mc_model}. 
Disk parameters: $M = 10^6 M_{\odot}$,
$\dot m = 0.3$, without an irradiation due to the warm absorber.}
\label{fig:pureCompton}
\end{figure}

Although the condition $A = 2$ roughly determines the slope of the power-law
part of the spectrum, the Monte Carlo computations make it possible to see some trends
with the change of optical depth. In Fig.~\ref{fig:slope} we show
the spectral slope determined by a linear fit to the resulting spectrum in 
two soft X-ray bands, for 0.5 keV to 2 keV and for 0.1 to 2.4 keV, for the
temperature of the input photons fixed at 5 eV. 
We see that generally the spectrum steepens with increasing 
 optical 
depth.  Analogous trends were already 
seen 
by Haardt \& Maraschi (1991) as well as for example by
Abrassart \& Czerny (2000) when they fixed the spectral slope but determined
the amplification factor as a function of the temperature of the medium.
The effect is more strongly visible for the slope measured in a 
narrower band since the spectrum displays some curvature due to the low 
value of the
electron temperature. Extreme examples of the Comptonized black body emission 
by optically thin and optically thick skin are shown in Fig.~\ref{fig:mc_model}.

For very low optical depth/high temperature
we see a considerable effect of the first scattering. This effect, well
known from Monte Carlo simulations, might in principle allow to determine
independently the optical depth of the Comptonizing medium. However, this
effect is not seen in the real data. This is not surprising - for example,
the effect is considerably reduced if we take the disk spectrum 
(instead of a single black body), as a source of the soft photons. A broader
input spectrum smears the first scattering feature, as we can see from direct
comparison of Fig.~\ref{fig:pureCompton} with Fig.~\ref{fig:mc_model}.

\begin{figure}
\epsfxsize=8.8cm \epsfbox{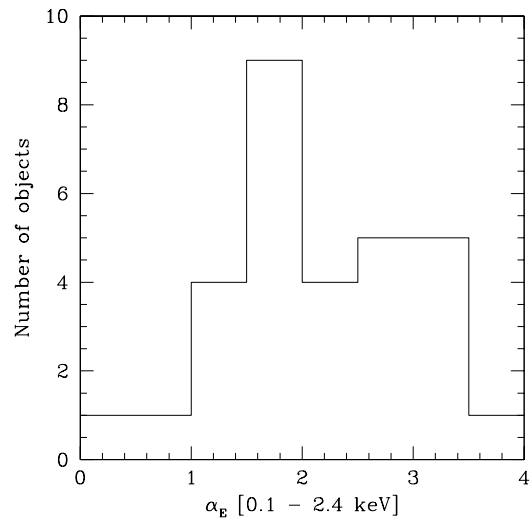}
\caption{The distribution of the X-ray slopes of NLS1 galaxies in the ROSAT 
sample from Boller et al. (1996).}
\label{fig:observ-slope}
\end{figure}

We can compare the expected range of soft X-ray slopes with the ROSAT
data for NLS1 galaxies (see Fig.~\ref{fig:observ-slope}).
High-accretion-rate objects tend to have a spectral slope of the order
of 1.7 (see Fig.~\ref{fig:examples} and Fig.~\ref{fig:observ-slope})
but with some dispersion around this value. Our model does not favour
any specific value of the temperature so the observed dispersion is
not surprising. Judging from the slope range (and accepting our
theoretical determination of the amplification factor) the optical
depth of the warm skin is typically between 1 and 10. This agrees with
the determination of the optical depth of the warm skin by detailed
modeling the X-ray data of PG1211+143 by Janiuk et al. (2000).

Computations of the radiation spectrum formed from the disk require
the determination of the disk fraction covered by the skin and subsequent
Comptonization of the locally emitted spectrum.  The extension of the
radiation-pressure-dominated zone was roughly estimated by Shakura \&
Sunyaev (1973) in their Eq.~(2.13). It depends only weakly on the black
hole mass and the viscosity coefficient. The extension of the
instability zone was computed more accurately by Janiuk \& Czerny
(2000), taking into account the disk vertical structure and a rather
accurate description of the opacity in the disk interior. The numerical
results preserve the scaling with accretion rate for large accretion
rates but give the extension of the unstable zone by a factor of 2
smaller than the simple Shakura \& Sunyaev formula. 

\subsection{Combined effect of warm skin and irradiation}

The fate of the radiation emitted by the disk and backscattered by the
warm absorber towards the disk surface depends on whether we deal
with the outer, bare part of the disk or with the inner part of the disk 
covered by the warm skin.

The part of the disk not covered by the warm skin is only weakly ionized
The incident flux, given by Eq.~(\ref{eq:irrad}) is efficiently thermalised
so finally the disk radiates locally as a black body emitter with the
effective temperature, $T_{eff}$, determined by the condition
\begin{equation}
   \sigma _B T^4_{eff} = F_{diss} + (1 - a)F_{irrad},
\end{equation}
where $F_{diss}$ is the energy dissipated viscously by the disk and 
described by the standard Shakura-Sunyaev formula. Here we assume that 
the albedo $a$ is equal to 0.2. 
This radiation is integrated over the disk surface.

The part of the disk covered by the warm skin is practically fully
ionized.  Photons from the optical/UV/soft X-ray band are scattered by
such a medium.  Those photons however also contribute to the cooling
by Comptonization. In principle we should include those photons as
entering the warm skin slab from above but this would complicate the
numerical scheme.  Therefore, at present, those photons are also
treated as a black body photons entering the slab from the bottom. A
future, more sophisticated approach should treat these photons more
carefully, in an iterative way.

Since Comptonization is a linear process with respect to the input photons, 
and the parameters of the warm skin do not depend in our model on the disk 
radius, the radiation going into the warm skin is first integrated over the 
disk surface, and next the spectrum of radiation leaving the warm slab is 
computed by a Monte Carlo simulation for such an incident flux.

The radiation from the outer and the inner disk part are summed. 
Finally, the radiation reaches the observer diluted by a factor 
$1 - \tau_{wa}$.

\subsection{Modeled exemplary spectra}

Since the aim of our model is to reproduce in a possibly natural way
the scaling of the spectra with mass we do not try to actually fit the
data points. Instead, we fix the model parameters in a possibly unique
way, following the presented model outline.

We fix the accretion rate at $\dot m = 0.3$,
appropriate for high-Eddington-ratio objects.

We fix the warm absorber properties assuming that the solid angle of
the warm absorber is equal to $0.5 \times 2 \pi$ 
and the optical depth of the warm
absorber for electron scattering is equal to 0.25 (i.e. the column
density is equal to $ 3.8 \times 10^{23} $ cm$^{-2}$.). We assume that
the warm absorber extends from 100 $R_{Schw}$ to $10 000 ~R_{Schw}$.

The extension of the warm skin covering the disk is determined by the
extension of the radiation pressure instability zone. 
For $\dot m = 0.3$, Janiuk \& Czerny (2000) give $r_{skin} = 300~ R_{Schw}$ 
and we adopt this value.

The Compton amplification is fixed at $w = 0.5$ which determines the relation
between the optical depth and the electron temperature. We choose
$T_{skin}= 3$ keV and $\tau_{skin} = 3.32$ 
but any combination from Fig.~\ref{fig:ampli} can be used.

\begin{figure}
\vskip -0.8 truecm
\epsfxsize=8.8cm \epsfbox{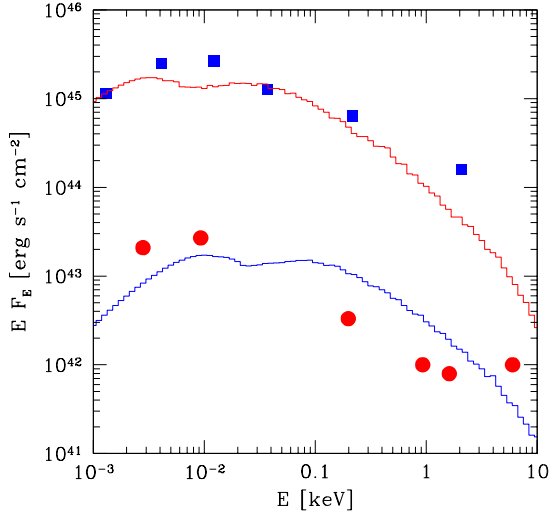}
\vskip -0.5 truecm
\caption{The exemplary accretion disk models (continuous histograms) 
for the composite spectrum of quasars (filled squares) and for Mrk 359
(filled circles) with irradiation and a warm skin covering the
radiation pressure dominated region. Model parameters are given in
Table~\ref{tab1}.}
\label{fig:best}
\vskip -0.2 truecm
\end{figure}

We now apply this model, without any specific adjustments of the
parameters, both to the low mass objects (Mrk 359, $M = 10^6 M_{\odot}$) 
and to the high mass composite spectrum ($M = 10^8 M_{\odot}$). The result 
is shown in Fig.~\ref{fig:best}, together with the data points representative 
for the low and the high mass objects.

\begin{figure}
\vskip -0.5 truecm
\epsfxsize=8.8cm \epsfbox{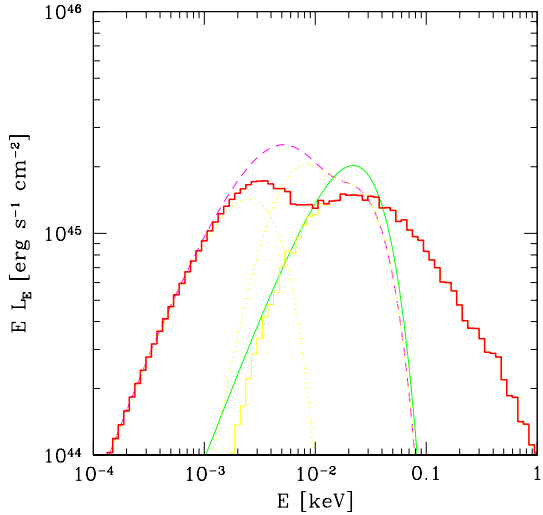}
\vskip -0.8 truecm
\caption{The standard disk model without the presence of the warm absorber and the warm skin (thin continuous line), disk affected by the warm absorber 
through irradiation and dilution (dashed line) divided into inner and
outer part (dotted lines), radiation of the inner disk modified by the
disk skin (continuous thin histogram) and the total spectrum (thick
continuous histogram) for a high mass object. Model parameters given
in Table~\ref{tab1}.}
\label{fig:decompo_laor}
\vskip -0.5 truecm
\end{figure}

The role of the irradiation and the Comptonization by the warm absorber is 
illustrated in Fig.~\ref{fig:decompo_laor} which shows the decomposition
of the exemplary final spectrum and separately the effect of the warm 
absorber scattering and the warm skin. The initial standard disk spectrum is
lowered down due to the presence of the warm absorber since only a fraction
$1 - \tau_{wa}$ of the radiation is transmitted to the observer. On the 
other hand the returning radiation enhances predominantly the optical part 
of the spectrum due to absorption and reemission of the incident flux by the
outer parts of the disk. 

The model provides a reasonable description of the data. Interestingly, 
it roughly reproduces the
normalization of the spectrum in the soft X-ray band without any specific 
adjustments of the model. The spectral shape does not show a strong 
dependence with the mass although some systematical trends are visible.
The spectrum is still systematically steeper in the optical/near-IR band
and the soft X-ray excess is more pronounced for low mass objects.

\begin{figure}
\vskip -0.5 truecm
\epsfxsize=8.8cm \epsfbox{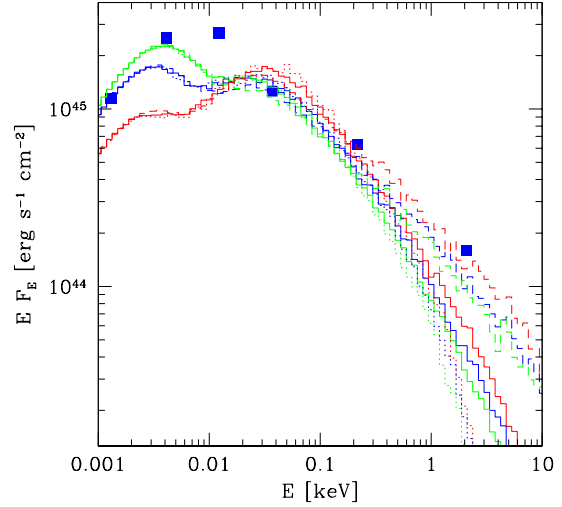}
\vskip -0.8 truecm
\caption{The set of 9 solutions obtained as a combination of the three
values of the warm absorber optical depth (0.1, 0.25 and 0.5) and
three values of the warm skin temperature (1, 3 and 30 keV, marked
with dotted, continuous and dashed histograms), with the optical
depths adjusted to the requested Compton amplification factor $A=2$
(i.e. $\tau_{skin}$ equal to 9.45, 3.32 and 0.548, correspondingly;
see Fig.~\ref{fig:ampli}), for $M = 10^8 M_{\odot}$. Other model
parameters as listed in Table~\ref{tab1}. Filled squares mark the
composite spectrum of quasars.}
\label{fig:various}
\vskip -0.5 truecm
\end{figure}

The arbitrariness in parameters still remaining in the model - neither
warm absorber column nor the warm skin temperature are specified - leads to
the dispersion in the predicted spectra. In Fig.~\ref{fig:various}
we show the family of 9 solutions determined as a combination of three values 
of the warm absorber optical depth and three values of the warm skin 
temperature. Solutions with lower skin temperature/optically thicker skin are 
more appropriate for softer spectra while those with higher skin 
temperature/thinner skin are appropriate for harder spectra. Higher warm
absorber optical depth increases the radiation flux in the optical/UV band.
Modelling any specific objects would require adjustment of the all model
parameters, including the mass and the accretion rate. However, the overall
dispersion among the solutions is not very large.  



\begin{table}
  \caption{Summary of model parameters.
  \label{tab1}}
  \begin{tabular}{lrr}
   \hline
\hline
Parameter & quasar composite &  Mrk 359 \\
\hline
$M [M_{\odot}]$ & $1.0 \times 10^8 M_{\odot}$ & $1.0 \times 10^6 M_{\odot}$ \\
\hline
$\dot m [L/L_{Edd}]$ &\multicolumn{2}{c}{~~~~~~~~~~0.3} \\
$r_{skin} [R_{Schw}]$ &\multicolumn{2}{c}{~~~~~~~~~~300} \\
$\tau_{skin}$ & \multicolumn{2}{c}{~~~~~~~~~~3.32} \\
$T_{skin}$ [keV] & \multicolumn{2}{c}{~~~~~~~~~~3}\\
$\tau_{wa}$ & \multicolumn{2}{c}{~~~~~~~~~~0.25} \\
$f_{wa}$ & \multicolumn{2}{c}{~~~~~~~~~~0.5}\\
$\delta$ &  \multicolumn{2}{c}{~~~~~~~~~~ 2} \\
$z_{min} [R_{Schw}] $ & \multicolumn{2}{c}{~~~~~~~~~~100} \\
$z_{max} [R_{Schw}] $ & \multicolumn{2}{c}{~~~~~~~~~~10 000}\\ 
\hline
 \end{tabular}
\end{table}

The scaling of the model with mass is not perfectly linear, as might have been
indicated by simple analytical arguments but it roughly holds, and the 
spectrum of the high mass object looks roughly as the spectrum of the low 
mass one, shifted up by a mass ratio in logarithmic scale. 

\section{Discussion}

The model presented in this paper applies to radio quiet AGN with relatively 
high Eddington ratio, like quasars and NLS1 galaxies. In these objects the
accretion disk basically extends down to the marginally stable orbit but the 
spectra show significant deviations from the simple prediction of a classical 
disk.

The model presented in this paper does not rely on many arbitrary parameters, 
as is usually the case for models supposed to reproduce precisely the 
observed spectra. The basic model parameters are just the black hole mass 
and the accretion rate. The inner part of the disk is covered by a warm skin,
and its radial extension is given by the 
condition of radiation pressure domination. The soft X-ray spectrum is
formed by Comptonization within this warm skin and the 
Compton amplification
factor is given by the stability criterion formulated in our paper. The 
optical depth and the warm skin temperature separately are not specified 
but the spectra
do not depend dramatically on the choice, as we can see from extreme cases
presented in Fig.~\ref{fig:mc_model}. Some dispersion of the spectral 
slopes is also seen in the data, which may indicate that there is indeed no
additional strong constraint on these two parameters and they may 
vary from source to source. The optical part of the spectrum is in turn 
affected by the back-scattering of a fraction of the photons towards the disk 
by the warm absorber. The warm absorber parameters currently are not as 
constrained by
physical considerations as the warm skin.

Still, the model is based on several assumptions which may, or may not be,
appropriate.

\subsection{Scaling of the magnetically transported flux towards warm skin}

Since the adopted (linear) dependence of the flux tube covering factor adopted 
in Eq.~(\ref{eq:mag_cooling}) may be questioned we also consider a more general
approach
\begin{equation}
W \propto B^2 v_A H^{\zeta},
\end{equation}
with $\zeta$ being an arbitrary coefficient.

In this case our previous stability criterion (see Eq.~(\ref{eq:stab_criterion}))
takes a more general form 
\begin{equation}
\beta > {2 - 2 \, w - 2 \zeta \, w \over 5 - 5 \, w - 1.5 \zeta \, w}.
\label{eq:stab_criterion2} 
\end{equation}
This means that the radiation pressure dominated disk (with $\beta \approx 0$) 
is stable if 
\begin{equation}
w = {1 \over 1 + \zeta}.
\end{equation}

The predicted Compton amplification factor of the warm skin is therefore
\begin{equation}
 A = {1 + \zeta \over \zeta}
\end{equation}
and the soft X-ray slope of the resulting spectrum gets steeper (softer) 
if the dependence of the covering factor on the disk thickness is stronger 
than linear ($\zeta > 1$). A considerably weaker dependence leads to very hard spectra and in
the limit of $\zeta \rightarrow 0$ all energy must be dissipated in the warm 
skin in order to stabilize the disk. In this limit our approach reduces to
the result obtained by Svensson \& Zdziarski (1994): if all energy is 
dissipated in the hot corona the underlying disk is dominated by the gas 
pressure, and therefore it is stable.

The value of $\zeta \approx 1$ seems to lead to the Compton amplification 
factor nicely reproducing the observed range of soft X-ray slopes in NLS1,
as seen from the comparison of Fig.~\ref{fig:slope} and 
Fig.~\ref{fig:observ-slope}. 
What is more, it explains why the disks seem relatively stable even if
dominated by the radiation pressure. 

\subsection{High and Very High State in our model}

Our model may also explain why the flow in
galactic sources is rather stable in the High State but becomes
unstable when the source enters the so called Very High State. 

The luminosity 
corresponding to the High State is roughly about 0.1 in Eddington units while
at the Very High State it is about 1. A standard disk dominated by the radiation
pressure is expected to be unstable at these luminosities and the 
stabilization is possible at luminosities by a factor of a few higher than the
Eddington limit due to the effect of advection (Abramowicz et al. 1988). 

The observed sources show the opposite behaviour.
The disk
component in the Soft State spectrum of Cyg X-1 was shown to be stable
(Churazov et al. 2001). A brighter source,
the microquasar GRS 1915+105 (which is radiating roughly at the 
Eddington limit, see e.g. Sobolewska \& \. Zycki 2003) 
frequently enters the extended state in which it shows
strong outbursts (e.g. Belloni et al. 2000 and references therein) 
lasting a few thousands of 
seconds, corresponding to the
viscous timescale in the inner disk region.

We can interpret this behaviour within the frame of our model. When the
source is below the Eddington limit, the presence of the warm skin 
stabilizes the disk. 
When the source approaches and/or exceeds the Eddington limit, the disk thickness
$H$ approaches the disk radius, $r$. In this case the surface covered by the
magnetic tubes approaches the whole surface and ceases to respond to the 
temperature changes. In this limit, the situation practically reduces to 
the case of $\zeta = 0$ and the radiation-pressure-dominated disk interior 
remains unstable, which leads to semi-periodic outbursts of a large amplitude,
as usually predicted by computations of the disk time evolution under 
radiation 
pressure instability (e.g. Szuszkiewicz \& Miller 1998, Szuszkiewicz \& Miller
2001, Janiuk et al. 2002). 

Of course we should keep in mind that there are two alternative 
explanations of the stability of the disk in the 
High State. The first one is that the assumption about the viscosity scaling 
with the total pressure is incorrect and should be modified 
(e.g. Milsom et al. 1994, Szuszkiewicz 1990). The second one is the 
possibility 
of an uncollimated outflow which also stabilizes the disk (Janiuk et al. 2002). 
Nevertheless, the warm skin offers an interesting possibility.

It is more difficult to give a similar discussion in the case of AGN due to 
intrinsically longer timescales and, consequently, poor knowledge of the 
evolutionary trends. Analogs of outbursts in GRS 1915+105 lasting 
$\sim 1000$ s would take $100 - 10^4$ yr in an AGN. However, 
large variations seen in NLS1 may be related to the discussed disk 
instability. This would mean that NLS1 are massive counterparts to galactic 
sources in their Very High States, as already argued by 
\Agata \& Czerny (2000).

\subsection{Optically thin or optically thick skin/corona?}

The recent Chandra and XMM observations show no or very weak
spectral features in the soft X-ray range in a number of NLS1 galaxies (e.g.
Ton S180, \Agata et al. 2003, Vaughan et al. 2002). Therefore, in those
sources the soft X-ray slope must be of Compton origin. However, as discussed
e.g. by Vaughan et al. (2002), the knowledge of the spectral slope does not allow
us to determine the optical depth {\it and} the temperature of the medium, and 
more generally, does not allow us to tell whether the medium is Compton-thick 
or Compton-thin. In the present paper we do not provide any constraints 
for the optical depth of the heated layer. Comparing the predicted range of 
soft spectral slopes with the observed slopes we may favour, a posteriori, an 
optically thick medium with a relatively low temperature. 

\subsection{Skin/corona or outflow?}

The location of the Comptonizing medium is also not known. In the present 
paper we assumed that this medium constituted a layer above the accretion 
disk. However, an alternative picture is also possible - that of an extended 
medium surrounding the nucleus, possibly in a form of an uncollimated outflow.

Such an outflow was considered by Janiuk et al. (2002) in the context of 
galactic sources and it was advocated by Pounds et al. (2003) and 
King \& Pounds (2003). The argument for the presence of an outflow comes
from the detection of narrow but blueshifted absorption lines in the soft X-ray 
spectrum of a Narrow-line quasar PG0844+349, indicating an outflow velocity of
order of 0.2 $c$. 

The bulk motion of the outflowing medium is high enough to cause significant
Comptonization of the scattering medium if the flow is optically thick (see
e.g. Janiuk et al. 2000). In this case the disk would be also stabilized,
as shown by Janiuk et al. 2002. 

The two models are difficult to distinguish at present. Perhaps the most 
promising approach would be through the variability studies.

\subsection{Irradiation by warm absorber}

In order to modify the spectral slope of the standard disk model 
in the optical band the irradiation must be considerable, i.e. the optical 
depth of the warm absorber for scattering should be at least of the order of 0.2.

The observations do not well constrain the total hydrogen column since the 
medium is highly ionized. For example, the model producing the soft X-ray 
tail of an AGN spectrum in an outflowing medium (Pounds et al. 2003) requires
an optical depth larger than 1 and King \& Pounds (2003) actually argue that 
a photosphere surrounds an active nucleus at a radius significantly larger 
than the Schwarzschild radius. A detailed spectroscopic analysis of more 
objects in the soft X-ray band may possibly provide stronger constraints.

\section{Conclusions}

We develop a model which is appropriate for high luminosity sources in terms
of the Eddington ratio, such as radio-quiet quasars and NLS1 galaxies. 

We propose two mechanisms - irradiation and marginal stability of the 
accretion disk warm skin - which together may be responsible for the 
relatively uniform spectral shape of those AGN. 

Irradiation results in softer optical/UV spectra than predicted by 
the classical disk 
models. The warm skin stabilizes the radiation-pressure-dominated disk and 
Comptonizes the disk emission thus producing the soft X-ray part of the 
spectrum. The condition of stability leads to a specific value of the
Compton amplification factor ($A=2$) which roughly determines the soft 
X-ray slope. Both effects lead to a radiation spectrum whose shape does not
strongly depend on the parameters.

Our results are in agreement
with the conclusion of Woo \& Urry (2002) 
that there is
no strong correlation between the Eddington ratio and the
mass. Masses of high-accretion-rate objects span a broad range, and 
a correspondingly broad range is covered by the bolometric luminosities.

Both mechanisms are studied at present in the simplified form and further
research is clearly needed to confirm their effectiveness and 
applicability for the modeling of the broad-band data. Future model 
should also account for the hard X-ray emission.

Lower-luminosity AGN (in terms of 
the Eddington ratio), like Seyfert galaxies,  are not expected to be 
represented by the discussed model. In those sources the disk evaporation is 
most probably very effective, as expected from the models 
(e.g. \Agata \& Czerny 2000, Liu et al. 2002), 
and the inner disk is replaced with a hot,
possibly advection dominated flow (e.g. Narayan \& Yi 1994). The small 
soft X-ray excess,
if present, is due to reprocessing of the X-ray emission by the outer disk
(e.g. Czerny \& \. Zycki 1994), and the iron line also formed in the outer disk
is mostly narrow. The skin-dominated part does not develop in those sources.

\begin{acknowledgements}

We thank Micha\l ~ Czerny, Joasia Kuraszkiewicz and Catherine Boisson 
for helpful discussions and the anonymous referee for very 
careful reading of the 
manuscript and for comments which helped to clarify several points.
Part of this work was supported by grants 2P03D00322 and PBZ-KBN-054/P03/2001
of the Polish
State Committee for Scientific Research, and by Jumelage/CNRS No. 16 
``Astronomie France/Pologne''.
\end{acknowledgements}

\end{document}